\documentclass[8pt,preprint]{aastex}
\usepackage{lipsum}
\usepackage{mathtools}
\usepackage{epstopdf}
\usepackage{cuted}
\usepackage[usenames, dvipsnames]{color}
\usepackage{hyperref}

\author{Anal Bhowmik \altaffilmark{1}, Narendra Nath Dutta \altaffilmark{2}, and Sourav Roy \altaffilmark{3} }
\email{nnd0004@auburn.edu, analbhowmik@phy.iitkgp.ernet.in}

\altaffiltext{1}{Department of Physics, Indian Institute of Technology Kharagpur, Kharagpur-721302, India.}
\altaffiltext{2}{Department of Chemistry and Biochemistry, Auburn University,  Alabama-36849, USA.}
\altaffiltext{3}{Department of Chemistry, Ben-Gurion University of the Negev, Beer-sheva-84105, Israel.}
\begin{document}


\title{Precise calculations of astrophysically important allowed and forbidden transitions  of Xe VIII}


\begin{abstract}
The present work reports transition line parameters for Xe VIII, which are potentially important for astrophysics in view of recent observations of multiply ionized xenon in hot white dwarfs. The relativistic coupled-cluster method is employed here to calculate the $E1$, $E2$, and $M1$ transition line parameters with high accuracy. The $E1$ oscillator strengths and  probabilities of $E2$ and $M1$ transitions are determined using theoretical amplitudes and experimental energy values. The calculated branching ratios and the lifetimes are supplemented to the transition parameters. Accurate presentation of these calculated data is crucial for density estimation in several stellar and inter-stellar media. 
\end{abstract}



\section{INTRODUCTION}

Precise estimations of the ultraviolet lines of xenon ions are important for  astronomy \citep{Dimitrijevic15}. There have been a number of recent theoretical and experimental endeavors to characterize the allowed and forbidden transition lines of different Xe ions \cite{Safronova2003, Glowacki2009, Migdalek2000, Ivanova2011, Saloman2004}. These lines can be used to diagnose densities and temperatures of various astronomical systems of interest.    Recently, Dimitrijevi\'{c} and coauthor \cite{Dimitrijevic15} have shown the importance of Stark broadening in spectral lines of Xe VIII in astronomical bodies, such as the extremely metal-poor halo PN H4-1 in primordial supernova \cite{Otsuka2013}. 

In experiments, transition probabilities are often determined from branching ratios, which are converted to an absolute scale using measured radiative lifetimes. These lifetimes depend on the allowed and forbidden  transition probabilities from the corresponding states. Though different many-body methods \cite{Safronova2003, Ivanova2011, Migdalek2000, Glowacki2009, Biemont2007} have been used to estimate the electric dipole transitions of Xe VIII over the years, there are discrepancies in published data. As a consequence,  there is a demand  to have correlation exhaustive relativistic  calculations for these transitions. It has been shown that accurate estimations of  transition matrix elements of Xe  ions can be used as indicators of accuracy in calculations of photoionization and electron scattering \cite{Muller2014}, and for plasma research. Experiments on electron-beam ion traps (EBIT)  motivate many theoretical studies of allowed and forbidden transitions of Xe$^{7+}$ \cite{Fahy2007}. Discrepancies between experimental measurements and theoretical calculations of photoionization spectra  \cite{Muller2014},  which may be due to population of important metastable states motivate us to do precise calculations of allowed and forbidden radiative transitions of Xe VIII.

In this present work, a non-linear version of the relativistic coupled-cluster theory (RCC)   based on the Dirac-Coulomb-Gaunt Hamiltonian, as discussed in our earlier papers \cite{Dixit2007, Dutta2016, Dutta2012a},  is used to calculate the transition amplitudes accurately. The RCC approach is an all-order extension of the limited-order many-body perturbation theory \cite{Coester1960, Mukherjee1975, Sahoo2004, Dutta2016}.  Also, compared to the well-known configuration interaction or multi-configuration
technique \cite{Szabo1996}, the presence of non-linear terms at a certain level of excitation makes RCC method more correlation exhaustive theoretically \cite{Dutta2011, Dutta2016}. 
 
In section 2, we will give a very brief picture of the theoretical tools used for the present calculations.   In section 3, we will present and discuss our calculated results and compare them with other works. In section 4, we will conclude our work with estimation of accuracy of the present results.

\section{THEORY}
According to the single-valence open shell coupled-cluster theory \cite{Morrison(1985), Lindgren(1987), Haque(1984), Dutta2012a}, the correlated wave function for an atom or ion of $N$ electrons, $\vert \Psi_v \rangle$, is related with the reference state wave function $\vert \Phi_v \rangle$ by the closed-shell cluster operator $T$ and the open-shell cluster operator $S_v$ following the relation:
\begin{equation}
\vert \Psi_v \rangle=e^T \{1+S_v\}\vert \Phi_v \rangle=e^T \{1+S_v\}a_v^\dagger\vert \Phi_0 \rangle.
\end{equation} 
Here `$v$' indicates the valence orbital occupied by a single electron. $\vert \Phi_v \rangle$ is  obtained by adding an electron (indicated by the creation operator $a_v^\dagger$) on top of the Dirac-Fock (DF)  ground state $\vert \Phi_0 \rangle$ of the closed-shell system (in this work the closed shell system is the ground state of Xe IX). All the core (fully occupied with electrons) and  virtual (completely unoccupied with electrons) orbitals with respect to the electronic configuration of this closed shell are generated at the DF level in the potential of $N-1$ electrons following the Koopman's theorem \cite{Szabo1996}. The operator $T$ includes all the single and double  excitations from the core orbitals. The operator $S_v$ includes also all the single and double excitations. However, in this operator a single excitation happens from the valence orbital `$v$'. In a double excitation corresponding to $S_v$, there is one excitation from the valence orbital and the other excitation from a core orbital  \cite{Roy2014}. In the present work, we limit our calculations to include single and double excitations.  This is a good approximation for calculations of atomic properties using the RCC approach as discussed in Ref. \cite{Dutta2016}. However, we also include a class of valence triple excitations in a peturbative way as mentioned in Ref. \cite{Majumder2007, Dixit2007}. The correlated single valence states generated by the coupled-cluster single, double and valence triple excitations are applied to calculate different electromagnetic transition properties of present interest \cite{Dixit2007, Dixit2007b} . 
 
The description of the oscillator strength (dimensionless) and transition probabilities (in s$^{-1}$) for the electric dipole ($E1$), electric quadrupole ($E2$) and magnetic dipole ($M1$) transitions are given elsewhere \cite{Dutta2011, Roy2014, Dixit2007b}.
The lifetime $\tau_{k\rightarrow i}$ of a transition can be calculated by considering all possible channels of emissions through $E1$, $E2$ and $M1$ transitions from the state $k$ to the state $i$,
\begin{equation}
\tau_{k\rightarrow i}=\frac{1}{ A_{k\rightarrow i}}.
\end{equation}
Here $A_{k\rightarrow i}$ represents the total transition probability (sum of the probabilities of the $E1$, $E2$, and $M1$ transitions) for the transition from $k$ to $i$. The branching ratio of any decay channel from a particular state depends on the decay probability of that channel compared to all the possible decay channels from that state to any lower energy state ($i$) and is defined as
\begin{equation}
\Gamma_{\text{Upper} \rightarrow \text{Lower}}=\frac{A_{\text{Upper} \rightarrow \text{Lower}}}{\sum_i A_{\text{Upper} \rightarrow i}}.
\end{equation}

\section{RESULTS AND DISCUSSIONS}
The DF orbitals  are  constructed here using the basis-set expansion technique \cite{MOTECC1990}. The radial part of each basis function is considered to have a Gaussian-type form \cite{MOTECC1990}. These Gaussian-type orbitals (GTOs) have two optimized parameters, $\alpha_0$ and $\beta$, as exponents. In the present calculations, these parameters are considered universal, i.e., the same values of $\alpha_0$ and $\beta$ apply to all the DF orbitals. For Xe VIII, these optimized universal parameters are derived to be 0.00525 and 2.73, respectively. The  optimization \cite{Roy2015} has been done with respect to the DF orbital wavefunctions obtained from the GRASP92  code \cite{Parpia2006}, which uses a precise numerical approach to solve the DF equations.  Here 33, 30, 28, 25, 21 and 20 GTO bases are considered for the DF orbitals of the $s$, $p$, $d$, $f$, $g$ and $h$ symmetries, respectively. The number of active orbitals considered for the abovementioned symmetries are 14, 14, 15, 14, 12 and 10 (including all the bound orbitals) at the RCC level of calculations.   These numbers of active orbitals for the different symmetries are chosen by observing a convergence of correlation contribution to the closed-shell energy with increasing number of these orbitals \cite{Dixit2008}.

In Table 1, we present ionization potentials (IPs) of the ground state and a few low-lying excited states in cm$^{-1}$. The maximum difference between our RCC calculations and the experimental results obtained from the National Institute of Standards and Technology (NIST) \cite{NIST2015} occurs in case of the $6s_{1/2}$ state and is around 0.9$\%$. For all the other states, the present IPs differ from the corresponding NIST values by less than around 0.3$\%$. The  fine-structure splittings (FSS) of the different terms as calculated by the RCC approach and their comparison with experimental data can be found also in Table 1. Because of high ionization, one may expect a substantial contribution from relativistic effects, even at the electron-electron interaction level. Therefore, for precise estimation of transition amplitudes, we need to consider leading order relativistic correction properly at this level. Our RCC calculations can take care of this by considering the unretarded Breit or Gaunt interaction, which we reported in our earlier calculations \cite{Dutta2012a, Roy2014, Roy2015, Dutta2016}.  Fig. 1. highlights the Coulomb correlation contributions (with the Dirac-Coulomb Hamiltonian)  and the Gaunt interaction effects (difference between the results obtained with the Dirac-Coulomb-Gaunt and the Dirac-Coulomb Hamiltonians) on the IPs. As expected \cite{Dutta2012a, Roy2014}, the correlation contributions to the spin-orbit multiplets of a particular symmetry are almost the same, whereas the Gaunt contributions deviate from this trend.  The correlation contributions span between 0.5$\%$ to 3$\%$ through all the states and results are consistent with the values as calculated using the third-order relativistic many-body perturbation theory (RMBPT(3)) \cite{Safronova2003}.  The Gaunt effect increases the IP values for most of the states and varies between -0.06$\%$ to 0.05$\%$ .  

Table 2 shows the transition amplitudes and oscillator strengths of $E1$ transitions in  both the length and velocity gauge forms \cite{Johnson2006, Grant2007}. The oscillator strengths are calculated using the present RCC amplitudes and the experimental wavelengths from the NIST. The use of experimental wavelengths with respect to the corresponding RCC wavelengths eliminates any kind of theoretical uncertainty in our estimated values that can arise from the wavelengths or excitation energies. The $E1$ lines,  presented in this table, fall in the ultraviolet region of the electromagnetic (EM) spectrum and are especially useful for space telescope based astronomy. There is a good agreement, indicating the accuracy of the calculations, between most of the results obtained from the length and velocity gauges apart from the $4f_{5/2} \rightarrow 5d_{3/2,5/2}$ and $4f_{7/2} \rightarrow 5d_{5/2}$ transitions. However, in these cases, the results from the length gauge are in good agreement with  other accurate results as seen in Table 2. This supports the previously observed  \cite{Grant2007} better stability of a many-body calculation for a transition using the length gauge form compared to the velocity gauge form.  Also, comparisons are made here for the other transitions with  the available theoretical results in the literature \cite{Safronova2003, Ivanova2011, Migdalek2000, Glowacki2009, Biemont2007}.  The present values show very good agreement with the RMBPT(3) values \cite{Safronova2003} where wavefunctions are calculated by treating the correlation up to the second order in the perturbation theory \cite{Blundell1987}. Though our RCC method is an all-order extension of RMBPT(3)  \cite{Morrison(1985), Sahoo2004}, it is expected that calculations up to this second order can account for the dominant portion of the all-order correlation corrections to the wavefunctions for $E1$ amplitudes \cite{Morrison(1985), Sahoo2004}. Our results are in good agreement with the configuration interaction  Dirac-Fock  values for $5s_{1/2}$ $\rightarrow$ $5p_{1/2,3/2}$ transitions as calculated by Glowacki and Migdalek \cite{Glowacki2009}. They did their calculations by using DF wavefunctions generated with  integer occupancy (CIDF) and non-integer occupancy  (CIDF($q$)) of outermost core orbital to characterize the core-valence correlation \cite{Glowacki2009}.  They considered the most important configurations of single and double excitations as well as few triple excitations to a few low-lying states. However, our calculation is extended to a larger active orbital space and therefore, handles a larger amount of excited configurations. The valence triple excitations are implemented in our calculations through a perturbative treatment which is mentioned in the Section: 2. In the past, a large number of oscillator strengths for Xe VIII were calculated by including core polarization (CP) correction only as a correlation contributing factor. Migdalek and Garmulewicz \cite{Migdalek2000} calculated the oscillator strengths for a few transitions of Xe VIII using two different types of approach. In both these approaches, they used a polarization potential between the core and the valence electron of the type $V_{P}(r)=-\frac{1}{2}\alpha r^2(r^2+r^2_0)^{-3}$ \cite{Migdalek2000}, where $\alpha$ is the dipole polarizability of the ionic core computed by Fraga et al. \cite{Fraga1976} and $r_0$ is a cut-off radius, which has been taken as the mean radius of the outermost core orbitals calculated at the DF level.  In their first approach, they used  the core polarization augmented DF method, which they defined as DF+CP \cite{Migdalek2000}. However, in their second approach, which they defined as DX+CP,  three different forms of local approximation for the valence-core exchange potential were used \cite{Migdalek2000} instead of the non-local exchange potential as used in the DF theory. These three forms employ semi-classical exchange, free-electron-gas approximation for exchange and classically approximated free-electron-gas exchange. They indicated these three forms of the model potentials by abbreviations SCE, HFEGE and CAFEGE, respectively \cite{Migdalek2000}.   The calculations of Bi\'{e}mont et al. \cite{Biemont2007} were performed using the relativistic Hartree-Fock theory with a core polarization correction (HFR+CP) using a semi-empirically fitted  polarization potential with the same form as mentioned above. Compared to these kinds of  treatments of the core polarization effect, the present RCC method is purely ab-initio in nature, and it calculates this effect in transition amplitudes using an all-order extension of the many-body perturbation theory \cite{Morrison(1985),  Sahoo2004} with a large number of active orbitals. In addition, we include a number of other correlation features including core correlation, pair correlations and other higher order effects \cite{Dutta2016}. Nevertheless, our RCC results show good agreement with most of the core polarization augmented DF or HFR results for the $5s_{1/2}$ $\rightarrow$ $5p_{1/2,3/2}$ and $5p_{1/2,3/2}$ $\rightarrow$ $5d_{3/2,5/2}$ transitions.  From our investigation,  we have found that these transitions are strongly dominated by the core polarization effect. Therefore, this kind of agreement is expected. There is a small discrepancy in the RCC and DF+CP oscillator strength values for the $5p_{3/2}$ $\rightarrow$ $5d_{3/2}$ transition  \cite{Migdalek2000}. However, the RCC oscillator strengths for the $5p_{1/2}$ $\rightarrow$ $5d_{3/2}$ and $5p_{3/2}$ $\rightarrow$ $5d_{5/2}$ transitions agree excellently with the corresponding DF+CP values. Here it should be mentioned that the allowed oscillator strengths in the transitions among the multiplets of $5p$ and $5d$ terms are expected to obey a certain ratio \cite{Cowan1982}. This is maintained well in the present RCC approach. Also, the HFR+CP value of the $5p_{1/2}$ $\rightarrow$ $6s_{1/2}$ transition \cite{Biemont2007} differs by a relatively large amount with respect to all other theoretical results including the RCC value. Here we should mention that in spite of the fact that our IP value for the $6s_{1/2}$ state differs by a relatively larger amount with respect to the corresponding NIST value, the $E1$ amplitudes associated with this state are good in agreement comparatively with the other accurate values. This may be due to an overall good quality of the $6s_{1/2}$ state wavefuncion in the region where the transitions associated with this state are peaked.   

In Table 3, the $E2$ and $M1$ transition amplitudes and probabilities  are presented with the corresponding experimental transition wavelengths. Here also, both types of transition rates are calculated using the RCC amplitudes and the experimental transition wavelengths. Comparison of the length and velocity gauge form of $E2$ transitions is also shown in the table. The gauge matching is very good  here, except the $4f_{7/2} \rightarrow 4f_{5/2}$ transition. However, for this particular transition, the large gauge disagreement happens at the DF level of calculation. We have also seen this disagreement between the length and velocity gauge values at the DF level as obtained from the GRASP 92 Code \cite{Parpia2006}. Therefore, this gauge discrepancy is not a demerit of our computational approach, but may be a consequence of a numerical problem in the velocity gauge calculations.   As seen from this table, $E2$ transition rates ($A^{\text{E2}}$)    are the strongest for the transitions $5f_{5/2}   \rightarrow 5p_{1/2}$, $5f_{7/2} \rightarrow 5p_{3/2}$, $5g_{7/2} \rightarrow 5d_{3/2}$, $5g_{9/2}   \rightarrow 5d_{5/2}$, $5d_{5/2}   \rightarrow 5s_{1/2}$ and $5d_{3/2}   \rightarrow 5s_{1/2}$. These UV forbidden emission lines have transition probabilities of the order of $10^6$ s$^{-1}$, which is around two order less than the average probability of allowed transitions. The fine structure transitions have a stronger $M1$ transition probability compared to the corresponding $E2$ transition probability. The multi-configuration Dirac Fock (MCDF) or multi-configuration Dirac Hartree Fock  (MCDHF) values of the $M1$ transition probability for the $4f_{7/2}\rightarrow 4f_{5/2}$ transition, as calculated by Grumer et al. \cite{Grumer et al. 2014} and Ding et al. \cite{Ding2012} in two separate works, are seen to agree excellently with the present result.  

In Table 4, the branching ratios ($\Gamma_{\text{RCC}}$) are calculated using Eq. (3).  Here we compare our data with the available theoretical results of  \cite{Ivanova2011}. Due to the negligible contributions from the $E2$ and $M1$ transitions compared to the $E1$ transition to the decay rates, we use probabilities of $E1$ transitions (in length gauge) to calculate $\Gamma_{\text{RCC}}$ approximately. The other available theoretical results were calculated using a relativistic perturbation theory with a zero-approximation model potential (RPTMP) \cite{Ivanova2011}. The discrepancies between the results obtained from the RPTMP
method  ($\Gamma_{\text{RPTMP}}$)  and the RCC theory are discussed in the next paragraph.

In Table 5, we compare the present lifetimes for few transitions of Xe VIII with the experimental results as well as few theoretical values. Bi\'{e}mont et al. calculated these lifetimes using the HFR+CP and MCDF methods \cite{Biemont2007}. They also mesaured these lifetimes using beam-foil spectroscopy (BFS) \cite{Biemont2007}. Except the $5d_{3/2}$ $\rightarrow$ $5p_{1/2}$ transition, the RPTMP lifetimes as calculated by Ivanova are seen to poorly agree with the corresponding theoretical results. This may be a consequence of the poor accuracy of her wavefunctions for some states due to numerical difficulties as pointed by her \cite{Ivanova2011}. Here it is to be noted that the RPTMP lifetime for $5d_{5/2}$ $\rightarrow$ $5p_{3/2}$ transition agrees excellently with the BFS measurement. However, as mentioned in the work of Bi\'{e}mont et al. \cite{Biemont2007}, the disagreement between their calculated theoretical lifetimes and the BFS measurement for this transition may be due to experimental difficulty. Therefore, this excellent agreement of the RPTMP lifetime seems to be fortuitous.  Nevertheless, the poor accuracy of the RPTMP wavefunctions is reflected also in the RPTMP wavelengths for the $4f_{5/2, 7/2}$ $\rightarrow$  $5d_{3/2, 5/2}$ transitions  (See Table 3 of  \cite{Ivanova2011}).  Values of these wavelengths are one order higher than the corresponding NIST and RCC wavelengths. Due to such type of difference in performance of the RCC and RPTMP theories, $\Gamma_{\text{RPTMP}}$ values differ significantly from the corresponding $\Gamma_{\text{RCC}}$ results in a few cases in Table 4.

\section{CONCLUSION}
We have presented the  amplitudes and oscillator strengths of allowed UV transitions in length and velocity gauges for the ion Xe VIII using a highly correlated relativistic many-body approach.  In addition, the amplitudes and probabilities of $E2$ transitions in both the gauges and $M1$ transitions in the length gauge for this ion have been estimated using the same approach.  The good agreement of our calculated ionization potential energies with the experimental measurements and the small differences between the length and velocity gauge values for most of the transition amplitudes indicate a high accuracy of our calculations. The merit of the present oscillator strengths are compared in detail with respect to the other theoretical values. The present branching ratios are highly accurate compared to the earlier results. A detailed comparison with other high-quality calculations indicates that our oscillator-strength values for the $E1$ transitions have uncertainties of about 5\% on average.  Similarly, the average gauge discrepancy for all the $E2$ transition probabilities (except $4f_{7/2}\rightarrow 4f_{5/2}$ transition) is around 8\% and this may be considered as the uncertainty in the probability values.  The spectroscopic data of the present work will be useful for the studies of different astrophysical systems including low density hot plasma. To the best of our knowledge, almost all the forbidden transition probabilities are reported here for the first time in the literature.

\begin{acknowledgments}
We are grateful to Professor B. P. Das and Professor R. K. Chaudhuri, Indian Institute of Astrophysics, Bangalore, India and Dr. B. K. Sahoo, Physical Research Laboratory, Ahmedabad, India for providing the RCC code. We are very much thankful to Dr. S. Majumder, Indian Institute of Technology Kharagpur,  India for valuable discussions.
\end{acknowledgments}

\clearpage

\begin{table}[ht]
\caption{Ionization potentials (IPs) and fine structure  splittings (FSS) (in cm${^{-1}}$) of Xe VIII and their comparison with  experimental (Exp) values.}
\centering
\begin{tabular}{c  c  c c  c  c  c}
\hline\hline 
\hspace{1cm} & \multicolumn{4}{c}{IPs\hspace{0.1cm}} & \multicolumn{2}{c}{FSS\hspace{0.1cm}}\\
State           & RCC     & &
Exp${^a}$          &  &
RCC          & {Exp}${^a}$ \\ 
\hline
$5s_{1/2}$ & 854995 & & 854769 &  &    & \\ 
$5p_{1/2}$ & 737059 & & 738302 &  &   & \\
$5p_{3/2}$ & 718263 & & 719717 &  &  18796 & 18585\\
$4f_{5/2}$ & 588730 & & 589608 &  &  & \\
$4f_{7/2}$ & 588088 & & 589058 &  & 642 & 550\\ 
$5d_{3/2}$ & 543506 & & 544881 &  &  & \\
$5d_{5/2}$ & 540549 & & 541953 &  & 2957 & 2928\\
$6s_{1/2}$ & 455364 & & 459272 &  &    & \\
$5f_{5/2}$ & 356376 & & 357190 &   &  &  \\
$5f_{7/2}$ & 355922 & & 356751 &   & 454 & 439 \\
$5g_{7/2}$ & 283609 & & 284501 &    &   & \\
$5g_{9/2}$ & 283617 & & 284501 &    & -8 & 0\\ [0.2ex]
\hline
\begin{small}
$^a$ Reference: \cite{NIST2015}
\end{small}
\\
\end{tabular}
\end{table}

\begin{figure}
\centering
\begin{tabular}{cc}
\includegraphics[trim={2cm 0.1cm 1cm 10cm},width=120mm]{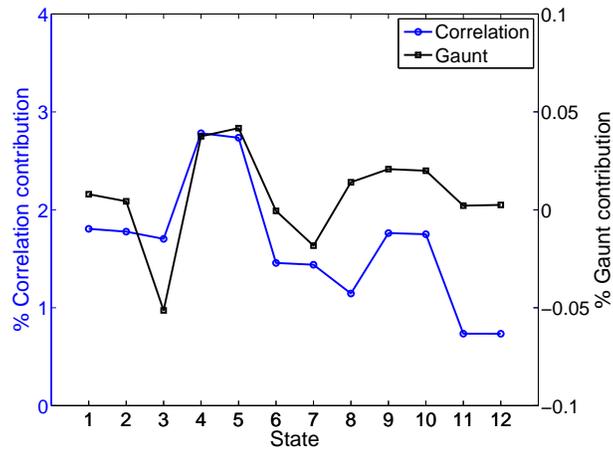}\\
\end{tabular}
\caption{Percentage of correlation and Gaunt contributions in the energy levels. Here numbers in the horizontal axis refer to the different energy states. They are 1$\rightarrow 5s_{1/2}$, 2$\rightarrow 5p_{1/2}$, 3$\rightarrow 5p_{3/2}$, 4$\rightarrow 4f_{5/2}$, 5$\rightarrow 4f_{7/2}$, 6$\rightarrow 5d_{3/2}$, 7$\rightarrow 5d_{5/2}$, 8$\rightarrow 6s_{1/2}$, 9$\rightarrow 5f_{5/2}$, 10$\rightarrow 5f_{7/2}$, 11$\rightarrow 5g_{7/2}$, 12$\rightarrow 5g_{9/2}$.}
\end{figure}

\begin{table}[ht]
\caption{ The RCC amplitudes ($O^{E1}_{\text{L}}$ \& $O^{E1}_{\text{V}}$) in atomic unit  and oscillator strengths ($f^{E1}_{\text{L}}$ \& $f^{E1}_{\text{V}}$) in atomic unit for length  and velocity gauges of $E1$ transitions along with the experimental wavelengths ($\lambda_{\text{Exp}}$) in {\AA}. The corresponding oscillator strengths obtained by the other works ($f^{E1}_{\text{Other}}$) are compared.}
\centering
\begin{tabular}{cccccccc}

\hline \hline

  \multicolumn{2}{c}{Levels\hspace{0.1cm}} & & & &  &  & \\

 Lower &  Upper &    $\lambda _{\text{Exp}}^*$        & $O^{E1}_{\text{L}}$ &  $O^{E1}_{\text{V}}$ & 
$f^{E1}_{\text{L}}$   & $f^{E1}_{\text{V}}$  &  $f^{E1}_{\text{Other}}$        \\ [0.2ex]
\hline
$5s_{1/2}  $  & $5p_{1/2}$     & 858.61  &   1.1736 &   1.1644  &   0.2436  &   0.2398 &  0.294$^a$, 0.234$^b$, 0.242$^c$, 0.253$^d$, \\ 
                      &                &        & 
         &        &         &       &    0.237$^e$,  0.237$^f$, 0.232$^g$, 0.223$^h$, 0.232$^i$
 
          \\
             &    $5p_{3/2}$    & 740.46    & 1.6705 &   1.6457  &   0.5724 &    0.5555  & 0.697$^a$, 0.550$^b$,  0.569$^c$, 0.596$^d$, \\ 
 
      &  & &   & &

 &   &  0.560$^e$, 0.563$^f$, 0.543$^g$, 0.522$^h$, 0.537$^i$
 \\

$5p_{1/2}  $ &    $5d_{3/2}$     & 517.01    & 1.8629 &    1.8279

  & 1.0195   

 & 0.9815
 &  1.189$^a$, 0.977$^b$, 1.020$^c$, 1.025$^d$, \\ 
 
      &  & &   & &

 &  &  1.003$^e$, 1.000$^f$, 1.057$^i$
 \\
 
  &   $6s_{1/2}$     & 358.38    & 0.6026   &   0.5767 

 & 0.1539 &    0.1409 & 0.160$^c$, 0.156$^d$, 0.155$^e$, 0.153$^f$, 0.199$^i$

   \\ 
 
 $5p_{3/2}  $ &    $5d_{3/2}$     & 571.96    & 0.8646   & 0.8496   

  & 0.0992 &     0.0958

 &  0.095$^b$, 0.089$^c$, 0.099$^d$, 0.097$^e$ \\
     &  & &   & &

 &  &  0.097$^f$, 0.095$^i$
  \\

   &   $5d_{5/2}$     & 562.54  &   2.5912    & 2.5432

  & 0.9064  
 & 0.8731 &  0.523$^a$, 0.868$^b$, 0.904$^c$, 0.907$^d$, \\ 
 
      &   &   &       & &

 &  &  0.889$^e$, 0.886$^f$, 0.875$^i$
 \\
    &    $6s_{1/2}$     & 383.96    & 0.9582 &   0.9121  

 &  0.1816   
 & 0.1645  & 0.188$^c$, 0.186$^d$, 0.184$^e$, 0.182$^f$, 0.186$^i$
\\

 $4f_{5/2}  $ &    $5d_{3/2}$     & 2235.79    & 1.5958   & 0.6806

  & 0.0577   
 & 0.0105 &  0.130$^a$, 0.058$^b$, 0.060$^i$
  \\
   
    &    $5d_{5/2}$     & 2098.42    &  0.4238  & 0.1998 

 & 0.0043  

 & 0.0010
 &  0.0044$^b$  \\
    
     &    $5g_{7/2}$    & 327.75    & 1.5869  & 1.5633  

  & 0.3890   & 0.3775

 &  0.3646$^b$, 0.354$^i$
  \\
 
 $4f_{7/2}  $ &    $5d_{5/2}$     & 2122.92  & 1.9075   &  0.8719  
 & 0.0651 &   0.0136
 &  0.075$^a$, 0.065$^b$, 0.068$^i$ \\
  
   &   $5g_{7/2}$     & 328.35  &   0.3073 &  0.3026   
  & 0.0109  
 & 0.0106
 &  0.0102$^b$  \\
  
    &   $5g_{9/2}$     & 328.35    & 1.8190  & 1.7919    & 0.3826   

 & 0.3713
 &  0.3595$^b$, 0.343$^i$
 \\

 $5d_{3/2}  $ &    $5f_{5/2}$    & 532.79    & 2.8328   &  2.7332  
 & 1.1437   
 & 1.0647

 & 1.099$^i$
  \\

 $5d_{5/2}  $ &    $5f_{5/2}$     & 541.23    &  0.7669    &  0.7410  
 & 0.0550   

 & 0.0514
 &  0.052$^i$
 \\
 
     &    $5f_{7/2}$     & 539.95  &   3.4223  & 3.3040 &   1.0981   

 & 1.0235
 &  1.032$^i$
 \\

 $5f_{5/2}  $ &    $5g_{7/2}$     & 1375.72    & 5.3626    & 5.4089  
 & 1.0583  

 & 1.0766
 & 1.071$^i$
  \\
 
  $5f_{7/2}  $ &    $5g_{7/2}$    & 1384.08   & 1.0330    &  1.0425  

  & 0.0293   
 & 0.0298
 &  0.030$^i$
  \\
  &    $5g_{9/2}$     & 1384.08    & 6.1114    & 6.1671

  & 1.0246  
 & 1.0434
 &  1.035$^i$
 \\
 
 \hline 
 
  \end{tabular}
 
 \begin{small}
\hspace{0.1cm}  $\lambda_{exp}^*$ are the Ritz wavelengths derived from experimental energy levels taken from  \cite{NIST2015}.
 \end{small}\\
 \begin{small}
\hspace{-10.9cm} $^a$ RPTMP method \cite{Ivanova2011}.
 \end{small}\\
 \begin{small}
 \hspace{-9.1cm} $^b$ RMBPT(3) method \cite{Safronova2003}.
 \end{small}\\
 \begin{small}
 \hspace{-8.0cm} $^c$ DF+CP method \cite{Migdalek2000}.
 \end{small}\\
 \begin{small}
 \hspace{-3.9cm} $^d$ DX+CP method with SCE model potential \cite{Migdalek2000}.
 \end{small}\\
 \begin{small}
 \hspace{-3.1cm} $^e$ DX+CP method with CAFEGE model potential \cite{Migdalek2000}.
 \end{small}\\
 \begin{small}
 \hspace{-3.5cm} $^f$ DX+CP method with HFEGE model potential \cite{Migdalek2000}.
 \end{small}\\
 \begin{small}
 \hspace{-4.3cm} $^g$ CIDF method with integer occupation number \cite{Glowacki2009}.
 \end{small}\\
 \begin{small}
 \hspace{-3.2cm} $^h$ CIDF($q$) method with non-integer occupation number \cite{Glowacki2009}.
 \end{small}\\
 \begin{small}
 \hspace{-9.8cm} $^i$ HFR+CP method \cite{Biemont2007}.
 \end{small}\\

\end{table}

\begin{table}[ht]

\caption{ Transition amplitudes ($O^{E2}$ ($O^{M1}$) ) and probabilities ($A^{E2}$ ($A^{M1}$)) of $E2$ ($M1$) transitions (in sec$^{-1}$) along with the experimental wavelengths ($\lambda_{\text{Exp}}$) in {\AA}. The notation $P(Q)$ in  case of transition probabilities  means $P\times 10^Q$. L and V represent the length and velocity gauge respectively.  Ritz wavelengths are used to calculate $A^{M1}_\text{Other}$.}
\centering 
\begin{tabular}{ccccccllll} 
\hline\hline
\multicolumn{2}{c}{Levels\hspace{0.1cm}} & & & &  &  & & &\\

  Lower     &   Upper    & $\lambda _{\text{Exp}}^ *$     & $O^{E2}_{\text{L}}$  
& $O^{E2}_{\text{V}}$ & \hspace{1cm}$A^{E2}_{\text{L}}$&  $A^{E2}_{\text{V}}$ & $O^{M1}$  &   $A^{M1}$  &  $A^{M1}_\text{Other}$ \\ [0.2ex] 
\hline 

$5s_{1/2}$  & $5d_{3/2}  $   & 322.70 & 2.8876 

  & 2.9149
& 6.6714(+05)& 6.7982(+05)
 &    && 

\\ 
 &   $5d_{5/2}$&   319.68

 & 3.5302 & 3.5446
 &  6.9674(+05) & 7.0242(+05)
 &    && 

 \\ 

$5p_{1/2}$   &  $5p_{3/2}  $   & 5380.68 & 3.1100 & 2.9785
 &

  6.0045(-01)&5.5074(-01)
 &  1.1519 & 5.7437(+01)& 
 \\ 
&   $4f_{5/2}$  & 672.52 & 2.5098 
  & 2.7071
 & 8.5467(+03) & 9.9433(+03)
 &    && 
 \\ 
 &   $5f_{5/2}$  & 262.39 & 3.3129 &3.3258
 &  1.6471(+06) & 1.6600(+06)
 &    & & 
\\ 
$5p_{3/2}$     & $4f_{5/2}$   & 768.59 & 1.3605 

  & 1.4889
& 1.2882(+03)& 1.5428(+03)
 &    && 
 \\
 &   $4f_{7/2}$  & 765.35 & 3.3511 & 3.6876
 &
 5.9866(+03)& 7.2493(+03)
 &     && \\ 
   
 &  $5f_{5/2}$    & 275.84 & 1.9247 & 1.9307
 &

   4.3300(+05)& 4.3570(+05)
  &    && 
 \\ 
 
 &   $5f_{7/2}$    & 275.51 & 4.6997 & 4.7052
 &
  1.9479(+06)& 1.9524(+06)
   &    & &  \\

 $4f_{5/2}  $    & $4f_{7/2}$    & 181818.18 & 0.8999 & 2.4123
 &

 5.7057(-10)& 4.1000(-09)
 &  1.8514 &  1.9228(-03)
 &\hspace{-0.4cm}1.9227(-03)$^a$\\
 
     &     &  & & 

 &

 & 
 &  &  
 &\hspace{-0.4cm}1.9277(-03)$^b$\\
   
 &  $5f_{5/2}$   & 430.26 & 2.1643 & 2.2797
&

 5.9296(+04)& 6.5788(+04)
 &  0.0299
  &5.0459(+01)
 & 
 \\
&  $5f_{7/2}$   & 429.45 & 0.8797 & 0.8969
&  

7.4168(+03) & 7.7096(+03)
&  0.0034
  & 4.9212(-01)
& 
 \\

$4f_{7/2}$    &   $5f_{5/2}$  & 431.28 & 0.8906  & 0.9093
&
   9.9224(+03)& 1.0343(+04)
 &  0.0141
  &1.1142(+01)
 & 
 \\
 & $5f_{7/2}$  &   430.46 & 2.5611 & 2.7021
&
  6.2129(+04) & 6.9159(+04)
 & 0.0612
   &1.5833(+02)
 & 
 \\

$5d_{3/2}$    &  $5d_{5/2}  $   & 34153.00 
& 3.0276  

  &  2.8753& 3.6821(-05)& 3.3210(-05)
  &  1.5488 &  2.7070(-01)& 
  \\
 
&  $6s_{1/2}$&  1168.10 & 4.5715 & 4.1909
 &

  5.3813(+03)& 4.5225(+03)
  &    & & 
  \\

&  $5g_{7/2}$   & 384.05 & 8.3453 & 8.4230
&

   1.1670(+06)& 1.1888(+06)
 &    & & 
  \\
$5d_{5/2}$     & $6s_{1/2}$  & 1209.46  & 5.6983 & 5.2459
 &

  7.0259(+03) & 5.9546(+03)
&    & & 
 \\
&  $5g_{7/2}$   & 388.42 & 2.8206 & 2.8465
&

   1.2597(+05)& 1.2830(+05)
 &    & & 
  \\
 
&   $5g_{9/2}$  & 388.42 & 9.9738 &  10.0619
&

  1.2601(+06)& 1.2825(+06)
 &    & & 
 \\

 $5f_{5/2}  $    &$5f_{7/2}$    & 227790.43 & 4.6121 & 4.3423
 &
 4.8554(-09)& 4.3040(-09)
  & 1.8515 &  9.7789(-04)& 
 \\  [0.2ex] 
\hline 
\end{tabular} 
 \begin{small}
  $\lambda_{exp}^*$ are the Ritz wavelengths derived from experimental energy levels taken from  \cite{NIST2015}.
 \end{small}\\
 \begin{small}
\hspace{-3.9cm} $^a$  Multi-configuration Dirac-Hartree-Fock calculations (MCDHF) \cite{Grumer et al. 2014}. 
 \end{small}\\
 
 \begin{small}
 \hspace{-7.0cm} $^b$ Multi-configuration Dirac-Fock (MCDF)  \cite{Ding2012}.
 \end{small}\\
\end{table}

 \begin{table}[ht]

\caption{ Branching ratios of different transitions and their comparison with the RPTMP values.}
\centering 
\begin{tabular}{ccc r}
\hline\hline

State(Upper)       &\hspace{-1.85cm}  State(Lower)         & $\Gamma_{\text{RCC}}$   & \hspace{1cm} $\Gamma_{\text{RPTMP}}$  \\ [0.5ex]              

\hline
$4f_{5/2}  $ &  \hspace{-2.85cm}  $5p_{1/2}$   & 0.86902 & \\

&   \hspace{-2.85cm} $5p_{3/2}$   & 0.13098 & \\

$5d_{3/2}  $ &   \hspace{-2.85cm} $5p_{1/2}$   & 0.85605 &  0.96236 \\

    &   \hspace{-2.85cm} $5p_{3/2}$   & 0.13618 & 0.03760 \\
    &   \hspace{-2.85cm} $4f_{5/2}$   & 0.00777 & 0.00003 \\

$5d_{5/2}  $ &   \hspace{-2.85cm} $5p_{3/2}$   & 0.98951 & 0.99985 \\ 
 
  &   \hspace{-2.85cm} $4f_{5/2}$   & 0.00051 & 0.00001 \\ 
   &   \hspace{-2.85cm} $4f_{7/2}$   & 0.00998 & 0.00014 \\

 $6s_{1/2}  $ &   \hspace{-2.85cm} $5p_{1/2}$   & 0.32724 &   \\ 
 
   &   \hspace{-2.85cm} $5p_{3/2}$   & 0.67276  &   \\ 
  
  $5f_{5/2}  $&   \hspace{-2.85cm} $5p_{1/2}$   & 0.00009 &   \\
  
    &   \hspace{-2.85cm} $5p_{3/2}$   & 0.00002 &   \\
    
 &   \hspace{-2.85cm} $5d_{3/2}$   & 0.93454 & 0.97001  \\
   
    &   \hspace{-2.85cm} $5d_{5/2}$   & 0.06535  &  0.02999 \\
    
   $5f_{7/2}  $&   \hspace{-2.85cm} $5p_{3/2}$   & 0.00010 &   \\  
   
   &   \hspace{-2.85cm} $5d_{5/2}$   & 0.99990  & \\ 
     
 $5g_{7/2}  $ &   \hspace{-2.85cm} $4f_{5/2}$   & 0.83513 &   \\
  
   &   \hspace{-2.85cm} $4f_{7/2}$   & 0.03114 &   \\
   
    &   \hspace{-2.85cm} $5d_{3/2}$   & 0.00005  & \\ 
    
   &   \hspace{-2.85cm} $5d_{5/2}$   & 0.00001  & \\ 
  
    &  \hspace{-2.85cm}  $5f_{5/2}$   & 0.12897 &  \\
   
    &   \hspace{-2.85cm} $5f_{7/2}$   & 0.00470 &    
      \\

 $5g_{9/2}  $ &   \hspace{-2.85cm} $4f_{7/2}$   & 0.86898 &   \\
  &   \hspace{-2.85cm} $5d_{5/2}$   & 0.00006  & \\
   
    &   \hspace{-2.85cm} $5f_{7/2}$   & 0.13096 &    
      \\[0.5ex]

 \hline 
 \begin{small}
 $\Gamma_{\text{RPTMP}}\Rightarrow$ Reference: \cite{Ivanova2011}
 \end{small}\\
\end{tabular} 
\end{table}

 \begin{table}[ht]

\caption{ Present lifetimes  ($\tau_{\text{RCC}}$) of the transitions in $10^{-9}$ s and their comparisons with the RPTMP ($\tau_{\text{RPTMP}}$), RMBPT(3) ($\tau_{\text{RMBPT(3)}}$), HFR+CP ($\tau_{\text{HFR+CP}}$), MCDF ($\tau_{\text{MCDF}}$), and experimental ($\tau_{\text{Exp(a)}}$ and $\tau_{\text{Exp(b)}}$ ) values. }
\centering 
\begin{tabular}{ccccccccc}
\hline \hline
\hspace{-4.85cm} Transitions  &  \hspace{-4.85cm} $\tau_{\text{RCC}}$ & $\tau_{\text{RPTMP}}$   & $\tau_{\text{RMBPT(3)}}$ & $\tau_{\text{HFR+CP}}$ & $\tau_{\text{MCDF}}$
& $\tau_{\text{Exp(a)}}$  & $\tau_{\text{Exp(b)}}$  \\  [0.5ex]          
\hline

\hspace{-4.75cm}$ 5p_{1/2}$ $\rightarrow$  $5s_{1/2}$    &  \hspace{-4.85cm} 0.45  & 0.37 & 0.47 &0.48&0.53
  & 0.52(3) & 0.50$\pm$0.05

\\ 
\hspace{-4.85cm} $5p_{3/2}$ $\rightarrow$ $5s_{1/2}$    &  \hspace{-4.85cm} 0.29  & 0.23

 & 0.30 &0.31&0.33 & 0.35(2) & 0.33$\pm$0.03 \\
\hspace{-4.85cm} $5d_{3/2}$ $\rightarrow$ $5p_{1/2}$    &  \hspace{-4.85cm} 0.08  & 0.07 
 
 &   &0.07&0.06 & 0.10(2) \\
\hspace{-4.85cm} $5d_{5/2}$ $\rightarrow$ $5p_{3/2}$    &  \hspace{-4.85cm} 0.08  & 0.14 

 &   &0.08&0.07 & 0.14(2) \\ [0.5ex]

\hline 
\begin{small}
\hspace{-2.85cm}$\tau_{\text{RPTMP}} \rightarrow$ Reference:\cite{Ivanova2011}.
\end{small}\\

\begin{small}
\hspace{-1.15cm}$\tau_{\text{RMBPT(3)}} \rightarrow$ Reference:\cite{Safronova2003}.
\end{small}\\

\begin{small}
\hspace{0.8cm}$\tau_{\text{HFR+CP}}, \tau_{\text{MCDF}}, \tau_{\text{Exp(a)}} \rightarrow$ Reference:\cite{Biemont2007}.
\end{small}\\

\begin{small}
\hspace{-1.95cm}$\tau_{\text{Exp(b)}} \rightarrow$ Reference: \cite{Cheng1979}.
\end{small}\\

\end{tabular} 
\end{table}

\clearpage


\begin{thebibliography}{}






\bibitem[Bi\'emont et al. 2007]{Biemont2007}
 Bi\'emont, \'E.,   Clar, M.,   Fivet, V., et al.    2007, EPJD, \textbf{44}, 23.
 \bibitem[Blundell et al. 1987]{Blundell1987}
Blundell, S. A., Guo, D. S., Johnson, W. R., \& Sapirstein, J., 1987, ADNDT, \textbf{37}, 103.


\bibitem [Cheng \&  Kim 1979]{Cheng1979}
 Cheng, K.-T., \&  Kim, Y.-K., 1979, JOSA, \textbf{69}, 125.
\bibitem[MOTECC-1990]{MOTECC1990}
 Clementi, E. (Ed.), {\it Modern Techniques in Computational Chemistry:
MOTECC-90}, (ESCOM Science Publishers B. V., 1990).
\bibitem [ Coester \&  Kummel 1960] {Coester1960}
 Coester, F., \&  Kummel, H., 1960, NucPh, \textbf{17}, 477.
 \bibitem[Cowan 1982]{Cowan1982}
Cowan, R. D.,  The Theory of Atomic Structure and Spectra, 1982, (Berkeley, CA: University of California Press) 






\bibitem[Dimitrijevi\'{c} et al. 2015]{Dimitrijevic15} Dimitrijev\'{c}, M. S.,   Simi\'{c}, Z.,  Kova\v cevi\'{c}, A.,   Valjarevi\'{c}, A., \&  Sahal-Br\'{e}chot, S., 2015, MNRAS, \textbf{454}, 1736.

\bibitem [Ding et al. 2012]{Ding2012} 
Ding, X. B.,  Koike, F.,  Murakami, I.,  et al.  2012, JPhB, \textbf{45}, 035003.
\bibitem [Dixit et al. 2008]{Dixit2008}
Dixit, G.,   Nataraj, H. S.,   Sahoo, B. K.,    Chaudhuri, R. K., \&  Majumder, S., 2008, PhRvA,     \textbf{77}, 012718.
\bibitem [Dixit et al. 2007a]{Dixit2007}
Dixit, G.,    Sahoo, B. K.,    Chaudhuri, R. K., \& Majumder, S., 2007, PhRvA,     \textbf{76}, 042505.
\bibitem [Dixit et al. 2007b]{Dixit2007b}
Dixit, G.,  Sahoo,  B. K.,  Deshmukh, P. C.,  Chaudhuri,  R. K., \&  Majumder,  S., 2007,  ApJS,        \textbf{172},  645.
 \bibitem [Dutta  \&  Majumder 2011]{Dutta2011}
 Dutta, N. N., \&  Majumder, S., 2011, ApJ,        {\bf 737}, 25.
 \bibitem [Dutta et al. 2012]{Dutta2012a}
 Dutta, N. N., \&  Majumder, S., 2012, PhRvA, {\bf 85}, 032512.
\bibitem [Dutta \&   Majumder 2016]{Dutta2016}
 Dutta, N. N., \&   Majumder, S., 2016,  InJPh,       {\bf 90}, 373.








\bibitem[Fahy et al. 2007]{Fahy2007}
Fahy, K.,   Sokell, E.,  O'Sullivan,  G.,   et al. 2007, PhRvA, \textbf{75}, 032520.
\bibitem[Fraga et al. 1976]{Fraga1976}
Fraga, S., Karwowski, J., \& Saxena, K. M. S., 1976, {\it Handbook of Atomic Data} (Amsterdam: Elsevier).







\bibitem [Glowacki  \&   Migdalek 2009]{Glowacki2009}
 Glowacki, L., \&   Migdalek, J., 2009,  PhRvA, \textbf{80}, 042505.
\bibitem [Grant 2007] {Grant2007}
Grant, I. P., {\it Relativistic Quantum Theory of Atoms and Molecules: Theory and
Computation}, (Berlin: Springer, 2007).
\bibitem [Grumer et al. 2014] {Grumer et al. 2014}
Grumer, J.,   Zhao, R.,  Brage, T.,   et al. 2014, PhRvA, \textbf{89}, 062511.


 
 
 
 
 
 
 
\bibitem [Haque  \&  Mukherjee 1984]{Haque(1984)}
 Haque, A., \&  Mukherjee, D., 1984, JChPh,       \textbf{80}, 5058.
 



\bibitem [Ivanova 2011]{Ivanova2011}
Ivanova, E. P., 2011, ADNDT, \textbf{97}, 1-22, and references therein.

\bibitem [Johnson 2006]{Johnson2006}
Johnson, W.  R.,  2006, {\it Lectures on Atomic Physics}.


\bibitem [Kaufman \& Sugar 1981]{Kaufman1981}
Kaufman, V. \& Sugar, J. 1981, PhyS, \textbf{24}, 738.
\bibitem[Kramida et al. 2015]{NIST2015}
 Kramida, A.,   Ralchenko, Y.,   Reader, J., \& NIST ASD Team (2015). NIST Atomic Spectra Database (ver. 5.3),  http://physics.nist.gov/asd [2015, December 27]. National Institute of Standards and Technology, Gaithersburg, MD.






\bibitem [Lindgren \&  Morrison 1985] {Morrison(1985)}
 Lindgren, I., \&   Morrison, J., in Atomic Many-body Theory, ed., Lambropoulos, G. E., and  Walther, H., 1985, (3rd. ed.; Berlin: Springer), 3.

\bibitem [Lindgren \&  Mukherjee 1987]{Lindgren(1987)}
 Lindgren, I., \&  Mukherjee, D., 1987, PhR, \textbf{151}, 93.
 
 
 
 
 

\bibitem[Majumder et al. 2007]{Majumder2007}
 Majumder, S.,  Sahoo, B.K.,   Chaudhuri, R.K.,   Das, B. P., \&  Mukherjee, D., 2007, EPJD, \textbf{41}, 441. 

\bibitem [Migdalek \&   Garmulewicz 2000 ]{Migdalek2000}
 Migdalek  J., \&  Garmulewicz, M., 2000, JPhB, \textbf{33}, 1735.

\bibitem [Mukherjee et al. 1975]{Mukherjee1975}
 Mukherjee, D.,   Moitra R. K., \&  Mukhopadhyay, A., 1975, MolPh,     \textbf{30}, 1861.
\bibitem [ M\"{u}ller et al. 2014]{Muller2014}
 M\"{u}ller, A.,   Schippers, S.,   Esteves-Macaluso, D.,    et al. 2014, JPhB, \textbf{47}, 215202.
 
 


\bibitem [Otsuka   \&  Tajitsu 2013]{Otsuka2013}
Otsuka, M.,  \& Tajitsu, A., 2013, ApJ, \textbf{778}  146. 






\bibitem [Parpia et al. 2006]{Parpia2006}
Parpia, F. A.,   Fischer, C. F., \&  Grant, I. P., 2006,  CoPhC,      {\bf 175}, 745.

 
 
 
 
 
 
 

\bibitem [Roy et al. 2014]{Roy2014}
 Roy, S.,  Dutta, N. N.,  \&  Majumder, S., 2014, PhRvA, \textbf{89}, 042511.
\bibitem [Roy \&  Majumder 2015]{Roy2015}
 Roy, S., \&    Majumder, S., 2015, PhRvA, \textbf{92}, 012508. 
 
 






\bibitem [Safronova et al. 2003]{Safronova2003}
 Safronova, U. I.,    Savukov, I. M.,    Safronova, M. S., \&  Johnson, W. R., 2003, PhRvA, \textbf{68}, 062505. 
 \bibitem [Sahoo et al. 2004]{Sahoo2004}
 Sahoo, B. K., Majumder S., Chaudhuri, R. K., Das, B. P., \& Mukherjee, D. 2004, JPhB, \textbf{37}, 3409. 
\bibitem[Saloman 2004]{Saloman2004}
 Saloman,  E. B., 2004, JPhCh,     \textbf{33}, 765.
\bibitem [Szabo et al. 1996]{Szabo1996}
\textit {Szabo, A.,  \&  Ostlund, N. S. 1996,  Modern Quantum Chemistry:
Introduction to Advanced Electronic Structure Theory} (Dover,Mineola). 
 
 
 
 

\end{thebibliography}
\end{document}